\documentstyle[aps,epsf,epsfig,twocolumn,multicol,amsmath]{revtex}  
\newcommand{\ee}{\end{equation}}
\newcommand{\pic}[2]{\epsfxsize #1 cm \epsffile{fig#2.eps} }
\newcommand{\be}[1]{\begin{equation}}
\newcommand{\bea}{\begin{eqnarray}}
\newcommand{\eea}{\end{eqnarray}} 
\newcommand{\eq}[1]{Eq.\ (\ref{#1})}
\newcommand{\fig}[1]{Fig.~\ref{#1}} 
\begin{document}  
\tighten   
%
\title{ \large\bf Semiclassical description of multiphoton processes}
\author{Gerd van de Sand$^{1}$ and Jan M. Rost$^{2}$} 
\address{$^{1}$ Theoretical Quantum Dynamics, 
Fakult\"at f\"ur Physik, Universit\"at Freiburg,\\
Hermann-Herder-Str.  3, D-79104 Freiburg,
Germany}
\address{$^{2}$ Max-Planck-Institute for the Physics of Complex Systems, 
N\"othnitzer Str. 38, D-01187 Dresden, Germany\\[3mm]
\rm (January 2000)}
\author{
\begin{minipage}{152mm}  
\vspace*{3mm}
We analyze strong field atomic dynamics semiclassically,
based on a full time-dependent description with the Hermann-Kluk propagator.
From the properties of the exact classical trajectories, in particular 
the accumulation of action in time, the prominent features of above
threshold ionization (ATI) and higher harmonic generation (HHG) are 
proven to be interference phenomena. They are reproduced quantitatively
in the semiclassical approximation. Moreover, the behavior of the action
of the classical trajectories supports the so called strong field approximation
which has been devised and postulated for strong field dynamics.
\draft
\end{minipage}
}
\maketitle

\section{Introduction}
In the last two decades multiphoton processes have been studied
intensively, experimentally as well as theoretically.  The inherent
time-dependent nature of an atomic or molecular excitation process
induced by a short laser pulse renders a theoretical description
problematic in two respects.  Firstly, a full quantum calculation in
three dimensions requires a large computational effort.  For this
reason, quantum calculations have been restricted to one active
electron in most cases \cite{KSK92b,PKK97}.  Secondly, an intuitive
understanding for an explicitly time dependent process seems to be
notoriously difficult, exemplified by pertinent discussions about
stabilization in  intense laser fields \cite{stabil,SJ93,stabilrecent}.    
Many studies have been carried out to gain an intuitive understanding
of the two most prominent strong field phenomena, namely High Harmonic
Generation (HHG) and Above Threshold Ionization (ATI).  In the well
established early analytical formulation by Keldysh, Faisal and Reiss
the atomic potential is treated as a perturbation for the motion of
the electron in a strong laser field \cite{KFR}.

This picture is still used in more recent
models, where the classical dynamics of the electron in the laser
field is explicitly considered, e.g.\ in Corkum's rescattering model
which can explain the cutoff observed in HHG for linearly polarized
laser light in one spatial dimension \cite{Cor93}.  
The corresponding
Hamiltonian reads \cite{PLKK9k}
\begin{equation} \label{laserallg}
H = H_0  + E_0f(t) \, x \, \sin(\omega_0 t + \delta)  \,,
\end{equation}
where $H_0 = \frac{1}{2}p^2 + V(x)$ is the atomic Hamiltonian, $f(t)$
is the time-profile of the laser pulse with maximum amplitude $E_0$,
and $\omega_0$ is the laser frequency.  The interaction of the
electron with the atom is specified by the potential $V$.

Lewenstein et al.\ extended Corkum's rescattering idea to a
quasiclassical model which contains one (relevant) bound state not
influenced by the laser field on the one hand side and electrons which
only feel the laser field on the other side \cite{Lew94}.  This simple
model explains qualitatively the features of HHG well.  The same is
also true for an alternative model, where the electron is bound by a
zero-range potential \cite{BLM94} However, the basic
question if and to which extent these multiphoton processes can be
understood semiclassically, i.e., by interference of classical
trajectories alone, remains unanswered.  It is astonishing that no
direct semiclassical investigation of the Hamiltonian \eq{laserallg}
has been performed while a number of classical as well as quantum
calculations for \eq{laserallg} have been published.  However, only
recently, a semiclassical propagation method has been formulated which
can be implemented with reasonable numerical effort.  This is very
important for the seemingly simple Hamiltonian \eq{laserallg} whose
classical dynamics is mixed and in some phase space regions highly
chaotic which requires efficient computation to achieve convergence. 
Equipped with these semiclassical tools we have studied multiphoton
phenomena semiclassically in the frame work of \eq{laserallg}.  In
comparison to the exact quantum solution we will work out those
features of the intense field dynamics that can be understood in terms
of interference of classical trajectories.
		
The plan of the paper is as follows. In section II we provide the tools 
for the calculation of a semiclassical, time-dependent wavefunction.
In section III we discuss Above-Threshold-Ionization
(ATI) and work out the classical quantities which structure semiclassically
the relevant observables. In section IV we use this knowledge for the
description of Higher-Harmonic-Generation (HHG).
Section V concludes the paper with a comparison of HHG and ATI from a
semiclassical perspective and a short summary.
\section{Calculation of the semiclassical wave function} 
A (multi-dimensional) wave function $\Psi_\beta({\bf x},t)$ can be expressed as
\begin{equation} \label{propgl}
\Psi({\bf x},t) \, = \, \int_{0}^t \! \! d{\bf x}' \, K({\bf x},{\bf x}',t) \, \Psi({\bf x}') \, .
\end{equation}
Here, $\Psi({\bf x}')$ is the initial wave function at $t=0$ and 
$K({\bf x},{\bf x}',t)$ denotes the propagator.
We will not use the 
well-known  semiclassical Van Vleck-Gutzwiller (VVG) propagator which is inconvenient
for several reasons. Firstly, one has to deal with caustics, i.e.\ singularities
of the propagator, and secondly, it is originally formulated
as a boundary value problem. For numerical applications much better suited
(and for analytical considerations not worse) is the so called  
Herman-Kluk (HK) propagator  which is a uniformized propagator in  
initial value representation  \cite{HK84,Kay94a}, formulated in phase space,
\begin{align} \label{HKP}
K^{\scriptscriptstyle{HK}}({\bf x},{\bf x}',t) & \, = \,\frac{1}{(2 \pi \hbar)^n}  \int \! \! \! \int \! \! 
d{\bf p} \, d{\bf q} \,\, C_{\bf q p}(t) \, e^{i S_{\bf q p}(t) 
/\hbar}\nonumber
\\
& \quad \quad g_\gamma({\bf x}; {\bf q}(t) , {\bf p}(t)) \,\, g^*_\gamma({\bf x}'; {\bf q}, {\bf p}) \, 
\end{align}
with
\begin{equation} \label{HKG}
g_\gamma({\bf x}; {\bf q},{\bf p}) \, = \, 
\left( \frac{\gamma}{\pi} \right)^{n/4}  
\exp \left( -\frac{\gamma}{2} \left( {\bf x} - {\bf q} \right)^2 + 
\frac{i}{\hbar} {\bf p} \left( {\bf x} - {\bf q} \right) \right) 
\end{equation}
and 
\begin{equation} \label{HKFAK}
C_{{\bf q p}}(t) \, = \, \left| \frac{1}{2} \left( 
{\bf Q_q} + {\bf P_p} - i \hbar \gamma {\bf Q_p} - \frac{1}{i \hbar \gamma} {\bf P_q} 
\right) \right|^{\frac{1}{2}} \, .
\end{equation} 
Each  phase space point (${\bf q},{\bf p}$) in the integrand
of \eq{HKP} is the starting point of a classical trajectory with action
 $S_{\bf q p}(t)$. The terms ${\bf X_y}$ in the weight factor $C_{\bf q p}(t)$
are the four elements of the monodromy matrix, ${\bf X_y} = \partial {\bf x}_t / \partial {\bf y}$.
The square root in \eq{HKFAK} has to be calculated in such a manner that 
$C_{\bf q p}(t)$ is a continuous function of $t$.
The integrand in \eq{HKP} is -- depending on the system -- highly oscillatory.
Although we restrict ourselves to one spatial dimension (see 
\eq{laserallg})
the number of trajectories necessary for numerical convergence 
can reach $10^7$.
We note in passing that an integration by stationary phase approximation over momentum and
coordinate variables reduces the HK-propagator to the VVG-propagator \cite{Gro99}.

In all calculations presented here we have  used  a Gaussian wave packet 
as initial wave function,
\begin{equation} \label{gaussallg}
\Psi_\beta(x') = \left( \frac{\beta}{\pi} \right)^{1/4} \exp \left( \frac{\beta}{2}  
\left( x' - q_\beta \right)^2 \right) \, .
\end{equation}  
With this choice, the overlap
\begin{equation}
f_{\gamma \beta} (q,p) \equiv \int \! g^*_\gamma({x}'; {q}, {p}) \, \Psi_\beta(x') \, dx' 
\end{equation}
can be calculated analytically and \eq{propgl} reads together with \eq{HKP} 
\begin{align} 
\Psi^{\scriptscriptstyle H \! K}_\beta(x,t) 
\, = \, & \left( \frac{4 \gamma \beta}{\alpha^2} \right)^{\frac{1}{4}}
\frac{1}{2 \pi \hbar} 
\int \! \! \! \int dp \, dq \,\, 
e^{i S_{q p}(t)/\hbar} \, \nonumber \\
& \quad C_{q p}(t) \, g_\gamma(x; q(t),p(t)) \, f_{\gamma \beta} (q,p) 
\label{hkpsi}
\end{align} 
with $\alpha = \gamma + \beta$. For all results presented here we have
taken $\gamma = \beta$.

For comparison with our semiclassical calculations we determined the 
quantum mechanical wave function using standard Fast Fourier Transform split operator methods \cite{FFT}.  \
\section{Above Threshold Ionization}
We start from \eq{laserallg} with $\delta = 0$ and use a rectangular
pulse shape $f(t)$ which lasts for $4.25$ optical cycles.  This
setting is very similar to the one used in\cite{JES88}.

The energy spectrum of the electrons 
can be expressed by the Fourier transform of the autocorrelation function
after the pulse, i.e. for times $t>t_f$,
\begin{equation} \label{ATI}
\sigma(\omega) \, = \, Re \, \, \int \limits_{t_f}^{\infty} \! e^{i \omega t} \, 
\langle \, \Psi(t) \, | \, \Psi_f \, \rangle \, \, dt \, ,   
\end{equation}
where $\Psi_f = \Psi(t_f)$ is the wave function after the pulse and 
correspondingly
\begin{equation}
| \, \Psi(t) \, \rangle  \, = \, e^{i H_0 (t-t_f)/ \hbar } \, | \, \Psi_f \, \rangle 
\end{equation}
is calculated by propagating $\Psi_f$ for some time with the atomic 
Hamiltonian $H_0$ only after the laser has been switched off.

\subsection{Quantum mechanical and semiclassical spectra for ATI} \label{secatirechnung}
We will present results for two types of potentials to elucidate the 
dependence of the semiclassical approximation on the form of the 
potential.
\subsubsection{Softcore potential}
\begin{figure}
\pic{8}{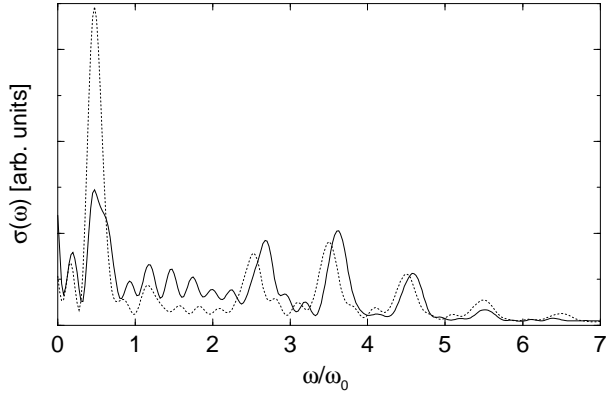}
\caption[]{Quantum mechanical (dotted line) and semiclassical (solid line) 
ATI spectra for the Hamiltonian of \eq{laserallg} with
$E_0=0.15$\! a.u., $\omega_0=0.148$\! a.u.\ and the softcore potential \eq{soft}.} \label{E015om0148}
\end{figure}

First we apply the widely used softcore potential \cite{JES88,LSW96}
\begin{equation} \label{soft}
V(x) \, = \, - \frac{1}{\sqrt{x^2 + a}}
\end{equation}
with $a=1$ and with an ionization potential $I_p = 0.670$\! a.u.. We 
have  checked that the correlation function differs little if calculated 
with the exact ground state or with the ground state wave function 
approximated by the Gaussian of \eq{gaussallg}
where $\beta = 0.431$\! a.u.\ and $q_\beta = 0$.
However, the semiclassical calculation is 
considerably simplified with a Gaussian as initial state as can be
seen from Eqs.~(\ref{gaussallg}-\ref{hkpsi}). Therefore we use 
this initial state and obtain the propagated semiclassical
wavefunction in the closed
form \eq{hkpsi}.
In \fig{E015om0148} the quantum and semiclassical results 
at a frequency $\omega_0 = 0.148$\! a.u.\ and a field strength $E_0 = 0.15$\! a.u.\ are 
compared. The Keldysh parameter has the value $1.14$.
The quantum mechanical calculation (dotted line) shows a typical ATI spectrum. 
Intensity maxima with a separation in energy of $\hbar \omega_0$ are clearly visible. 
The first maximum has the highest intensity while the second maximum is 
suppressed.
The semiclassical result  (solid line) is ambiguous: 
On the one hand there are clear ATI maxima with a separation of  $\hbar \omega_0$.
All peaks but the first one have roughly the  correct magnitude. 
Again the second maximum is missing.
On the other hand we see a constant shift 
(about $0.02$\!  a.u.) 
of the spectrum towards higher energies.  Therefore, a quantitative
semiclassical description is impossible, at least with the present
parameters and the softcore potential.  Next,  we will
clarify whether the shift in the spectrum is an inherent problem of a
semiclassical ATI calculation or if it can be attributed to properties
of the softcore potential.  
\subsubsection{Gaussian potential}
To this end we take a potential which has been used to model the 
``single bound state'' situation mentioned in the introduction 
\cite{Carla}. It is of Gaussian form
\begin{equation} \label{gausspotnochmal}
V(x) \, = \, -V_0 \, \exp \left( - \sigma x^2 \right) \, .
\end{equation}

With our choice of parameters $V_0 = 0.6$\! a.u.\ and $\sigma = 0.025$\! a.u.,
the potential contains six bound states and can be
 approximated, at least in the lower energy part, by a 
harmonic potential for which semiclassical calculations are exact.
Hence, the semiclassical ATI spectrum with this potential should be 
more accurate {\em if} the discrepancies in Fig.~\ref{E015om0148}
are due to the potential and not due to the laser interaction. 
The ground state wave function itself is again 
well approximated by the Gaussian \eq{gaussallg}  with $\beta = 0.154$\! a.u.\ and
$q_\beta = 0$.
The laser has  a frequency $\omega_0 = 0.09$\! a.u., 
a field strength $E_0 = 0.049$\! a.u., and a pulse duration of $4.25$ cycles.
The Keldysh parameter has the value $1.87$.

\begin{figure}
\pic{8}{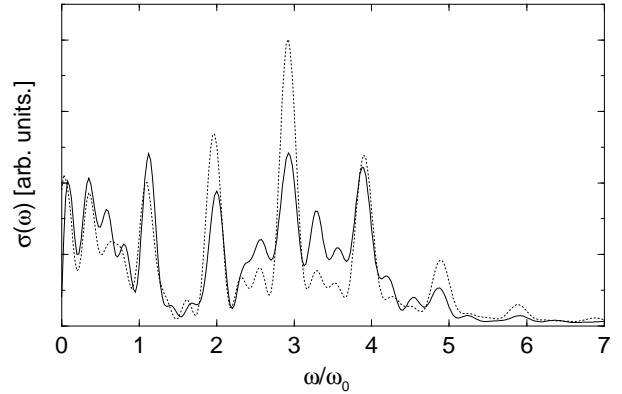}
\caption[]{Quantum mechanical (dotted line) and semiclassical (solid line) ATI spectra 
for the Hamiltonian of \eq{laserallg}  with
$E_0=0.049$\! a.u., $\omega_0=0.09$\! a.u.\ and the 
Gaussian potential \eq{gausspotnochmal}.} 
\label{E0049om009qmscl}
\end{figure}
We obtain a quantum mechanical ATI spectrum (dotted line in
\fig{E0049om009qmscl}) with six distinct maxima.  The semiclassical
spectrum (solid line) is not shifted, the location of the maxima
agrees with quantum mechanics.  Hence, one can conclude that the
softcore potential is responsible for the shift.  
 The height of the third maximum is clearly
underestimated and the details of the spectrum are exaggerated by the
semiclassical calculation.  Apart from these deviations the agreement
is good enough to use this type of calculation as a basis for a
semiclassical understanding of ATI.

\subsection{Semiclassical interpretation of the ATI spectrum} \label{secatiinterpret}
\subsubsection{Classification and coherence of trajectories}
With the chosen parameters most of the trajectories ionize during the pulse ($\sim 92 \, \%$).
We consider a trajectory as ionized if the energy of the atom 
\begin{equation}\label{timenergy}
\varepsilon(t)  = p(t)^2/2 + V(q(t))
\end{equation}
becomes positive at 
some time $t_{n}$ and remains positive, i.e. $\varepsilon(t) >0$ for
$t>t_{n}$. Typically, 
the trajectories ionize around an extremum of the laser field. Tunnelling can not be very important, 
otherwise the agreement between quantum mechanics and semiclassics would be much worse.
The Keldysh parameter of $1.87$ suggests that we are in between the tunnelling and 
the multiphoton regime. Interestingly,  the semiclassical description 
is successful  although we are  way below energies of 
the classically allowed over the barrier regime.

\begin{figure}
\pic{8}{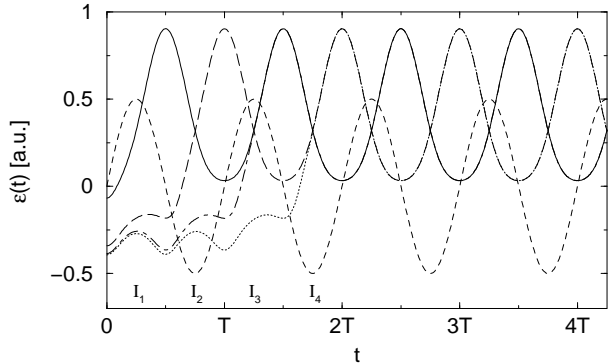}
\caption[]{Energy $\varepsilon(t)$ from \eq{timenergy} 
for trajectories ionized in the intervals $I_1$ (solid line), $I_2$
(dashed line), $I_3$ (dashed-dotted line) and $I_4$ (dotted line),
respectively.  For comparison, the laser field is plotted in arbitrary
units (thick dashed line).} \label{int1_4_et_pt}
\end{figure}
An obvious criterion for the classification of the trajectories is the
time interval of the laser cycle into which their individual ionization
time $t_{n}$ falls, see \fig{int1_4_et_pt}.  Typically ionization of trajectory happens around
$t_n = (2n-1)T/4$ when the force induced by the laser reaches a
maximum.  Hence, the ionized trajectories can be attached to time
intervals $I_n = [(n-1) T/2, n \, T/2]$.  In \fig{int1_4_et_pt} we
have plotted four trajectories from the intervals $I_1$ to $I_4$ which
end up with an energy $E = 0.36$\! a.u..   After ionization each
trajectory shows  a quiver motion around a mean momentum $p_f$   \cite{rem}.  
One can distinguish two groups of intervals, namely those with
trajectories  ionized with positive momentum $p_f$ (the
intervals $I_{2k-1}$) and those with trajectories with
negative $p_f$ (the intervals $I_{2k}$). 
These two groups contribute separately and incoherently to the energy spectrum
as one might expect since the electrons are easily distinguishable.
One can see this directly from the definition \eq{ATI} of the electron
energy spectrum.
For relative high energies $\hbar \omega$ the (short-range) potential may be 
neglected in the Hamiltonian $H_{0}$ and we get
\begin{align}
\sigma(\omega) \, & = \, Re \, \, \int \limits_{t_f}^{\infty} \! e^{i \omega t} \, 
\langle \, \Psi_f \, | \, e^{-i H_0 (t-t_f)} \, | \, \Psi_f \, \rangle \,\,  dt \, 
\nonumber \\ 
& \approx \, Re \, \, \int \limits_{0}^{\infty} \! e^{ i \omega t} \, \langle \, \Psi_f \, |  
\, e^{- i p^2 t / 2 \hbar} \, | \, \Psi_f \, \rangle \,\,  dt 
\nonumber \\ 
& \, = \, \, \int \limits_{-\infty}^{\infty} \!  
\delta \left( \omega - p^2/2 \hbar \right) \, 
\left | \Psi_f(p) \right | ^2 \, \, dp
\nonumber \\ 
& \, = \, \left(\left | \Psi_f(-\sqrt{2 \hbar \omega}) \right | ^2 + \left | 
\Psi_f(\sqrt{2 \hbar \omega}) \right | ^2 \right)
(\hbar / 2\omega )^{1/2} 
\nonumber \\[0.2cm] 
& \, \equiv \, \sigma_{-}(\omega) + \sigma_{+}(\omega) \, . \label{sigpm}
\end{align}
Hence, to this approximation, the ATI spectrum is indeed given by the
incoherent sum of two terms belonging to different signs of the
momenta of electrons ionized in different time intervals as described above.

\begin{figure}
\pic{8}{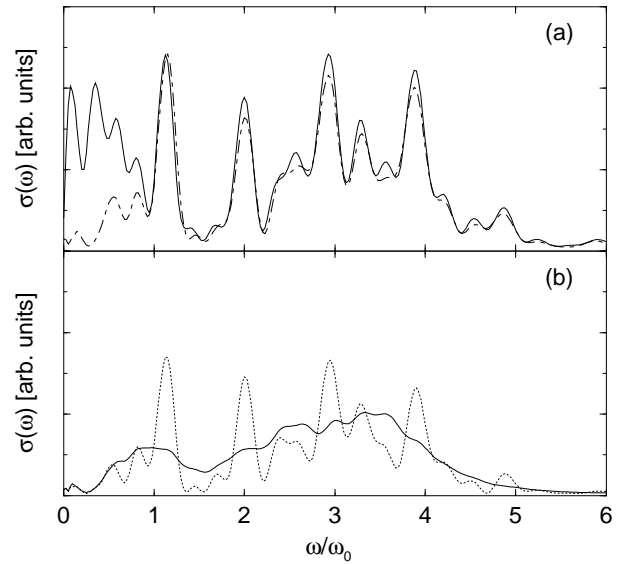}
\caption[]{Upper panel (a): Semiclassical spectrum as an incoherent sum 
$\sigma_+(\omega) + \sigma_-(\omega)$ (dashed-dotted line) compared with the full semiclassical
spectrum (solid line).
Lower panel (b): Semiclassical spectrum 
$\sigma_+(\omega)$ , 
constructed with trajectories from the intervals
$I_2$, $I_4$, $I_6$ and $I_8$ (dotted) compared to the incoherent sum 
$\tilde\sigma_{+}$
of spectra  that belong to the intervals $I_2$ to $I_8$ (solid line).} 
\label{kohsum}
\end{figure}

Figure  \fig{kohsum}(a) shows that \eq{sigpm} is a good approximation. 



Only for small $\omega$  the spectra do not agree,
where the kinetic energy is comparable with the (neglected) potential energy.

Quantum mechanically, all contributions from
 trajectories which lead to the same momentum $p_f$ of the 
electron are indistinguishable and must be summed  coherently. 
To double check that the interference
from different intervals $I_n$ is responsible for the
ATI peaks, we can artificially create a spectrum by an {\em
incoherent} superposition $\tilde\sigma_{+} = \sigma_{2} + \sigma_{4} +
\sigma_{6} + \sigma_{8}$ of contributions from trajectories ionized in
the intervals $I_{2j}$.  
This artificially 
incoherent sum (\fig{kohsum}(b)) shows  similarity neither
with $\sigma_{+}(\omega)$ nor with any kind of ATI spectrum.    

\subsubsection{Classical signature of bound and continuum motion in the laser field} \label{secsemmodati}
\begin{figure}
\pic{8}{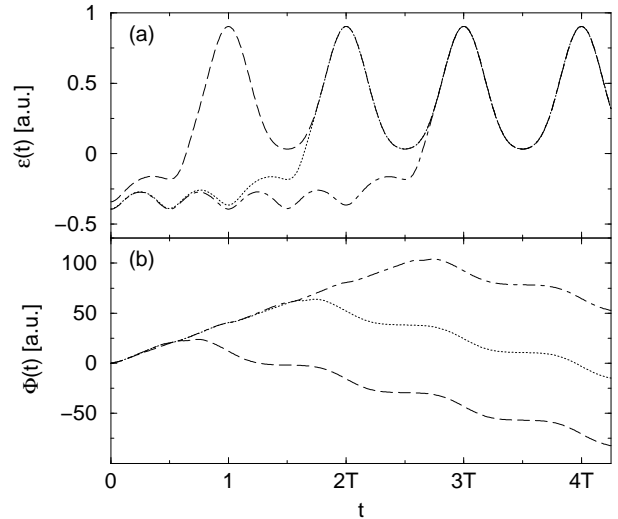}
\caption[]{Part (a) shows the atomic  energy $\varepsilon=p^2/2 + V(q)$ as 
a function of time for three trajectories from the intervals $I_2$ (dashed line), 
$I_4$ (dotted line) and $I_6$ (dashed-dotted line), part  (b) shows 
the corresponding phases $\Phi(t)$.}
\label{int246}
\end{figure}

The great advantage of an ab initio semiclassical description lies in the
possibility to make dynamical behavior transparent based on classical
trajectories, particularly in the
case of explicit time dependent problems where our intuition is not as
well trained as in the case of conservative Hamiltonian systems.
The classical quantities enter semiclassically mostly through the phase
factor
\begin{equation}\label{phidef}
 \exp \left(i [ S_{qp}(t) - p(t) q(t)]/\hbar\right) \equiv \exp[i\Phi/\hbar]
\end{equation}
which each trajectory contributes to the 
wave function \eq{hkpsi}. Although the prefactor 
$C_{qp}(t)$ in \eq{hkpsi} may be complex itself,
the major contribution to the phase comes from the effective action $\Phi$ in the
exponent of \eq{phidef}. 
Figure \ref{int246} shows the energy $\varepsilon$ of the atom and the accumulated
phase $\Phi$.
One can recognize a clear distinction between a quasi-free oscillation in the laser field
after the ionization and the quasi-bound motion in the potential.
The latter is characterized by an almost constant averaged bound energy
$\langle \varepsilon(t)\rangle$ (\fig{int246}(a)) of the individual trajectory
giving rise to an averaged linear increase of the phase (\fig{int246}(b)).
After ionization the phase decreases linearly with 
an oscillatory modulation superimposed by the laser field. 
The almost linear increase of $\Phi$ without strong modulation of the laser field
during the bound motion of the electron is remarkable, particularly looking
at the laser induced modulations of the bound energy seen in \fig{int246}(a).
The averaged slope of the phase (positive for bound motion, negative for
continuum motion) corresponds via $d\Phi/dt = -E$ to an averaged energy.
The behavior
can be understood by a closer inspection of the action
\bea\label{phase1}
\Phi(t)& \equiv& S_{qp}(t) - p(t) q(t)\nonumber\\
& =& \int_0^t(2T-H-\dot p(\tau)q(\tau) - \dot q(\tau)p(\tau))d\tau - qp\,.
\eea
Here, $T= p^2(t)/2$ refers to the kinetic energy and $H$ to the entire Hamiltonian
of \eq{laserallg}, the dot indicates a derivative with respect to time, and
$q \equiv q(t=0)$.
 With the help of Hamilton's equations and a little algebra $\Phi$ from
\eq{phase1} can be simplified to 
\begin{equation}\label{phase2}
\Phi(t) = -\int_0^t \left(\varepsilon(\tau)
- q(\tau)\frac{dV}{dq}\right)d\tau
\end{equation}
where $\varepsilon$ is the atomic energy \eq{timenergy}.
With \eq{phase2} we can quantitatively explain the slope of $\Phi$
in \fig{int246}(b). For the low energies considered the potential
\eq{gausspotnochmal} can be approximated harmonically,
\begin{equation}\label{parab}
V(q) \approx -V_0 + V_0\sigma q^2
\end{equation}
Averaging $\Phi$ over some time yields then $\Phi(t) \approx V_0t$,
for {\it any} bound energy of a classical trajectory since 
for an oscillator averaged kinetic and potential energy are equal.
Indeed, the numerical value for the positive slope in \fig{int246}(b) is
$0.6$\! a.u.\ in agreement with the value for $V_0$.

For the ionized part of the trajectories we may assume that the potential
vanishes. The corresponding solutions for  electron momentum $p(t)$ follows
 directly 
from Hamilton's equation $\dot p = -E_0\sin\omega_0t$,
\begin{equation} \label{qt}
 p(t) =   \frac{E_0}{\omega_0} \cos(\omega_0 t) +p,
\end{equation}
where $p$ is the mean momentum. 
Without potential the phase from \eq{phase2} reduces to 
$\Phi(t) = -\int p^2(\tau)/2 \, d\tau$ and we obtain with \eq{qt}
\bea
&& \Phi_c(t) \nonumber \\
&&  = - \frac{U_p}{2 \omega_0} \sin(2 \omega_0 t) -
\frac{E_0p}{\omega_0^2}\sin\omega_0t-(U_p + p^2/2)\, t   
\label{freewirkneu}
\eea
with the ponderomotive potential $U_p = E_0^2 / 4\omega_0^2$. We note in passing that
 \eq{freewirkneu} is identical to the time dependent phase
in the Volkov state (see the appendix).

\subsubsection{Semiclassical model for ATI}
The clear distinction between classical bound and continuum motion in the laser
field as demonstrated by \fig{int246} and illuminated in the last section,
allows one to derive easily the peak positions of the ATI spectrum. 
Moreover, this distinction also supports the so 
called strong field  approximation (e.g.\ \cite{Lew94,Lew95}) where electron dynamics in the laser field is
modelled by one bound state and  the
continuum. While this is postulated in \cite{Lew94} as an approximation
and justified a posteriori by the results the corresponding approximation
is suggested in the present context of a semiclassical analysis by the
full classical dynamics, i.e., the behavior of the trajectories, as shown in
\ref{int246}.  There, we have seen that each classical
bound motion leads to the characteristic linear increase of the phase.
If the entire phase space corresponding to the initial (ground state) wave
function is probed with many trajectories of different energy, the dominant
contribution will appear at the bound state energy which implies
 \begin{equation}\label{baction}
\Phi_b(t) \, \approx \, I_p \, t \,,
\end{equation}
where $I_p$ is the ionization potential. The time for which a trajectory does not
fall into one of the two classes, bound or continuum, is very short
(\fig{int246}).
Hence, we can approximately compose the true phase $\Phi = \Phi_b + \Phi_c$.
However, we don't know for an electron with  mean momentum $p$
when it was ionized. Hence, we have to sum over all trajectories 
with different ionization times $\tau$ 
but equal final momentum $p = p_f$ which leads
to the propagated wavefunction 
\bea\label{propa}
\Psi_f(t,p) \sim&& \int_{t_0}^td\tau
\exp[i/\hbar(\Phi_b(\tau)+\Phi_c(t)-\Phi_c(\tau))]\nonumber\\
\sim
&&\sum_{n,m}J_n\left(\frac{E_0p}{\omega_0^2}\right)J_m\left(\frac{U_p}{2\omega_0}\right)
\int_{t_0}^td\tau e^{i\tau\Delta_{mn}/\hbar}\,,
\eea
where the phase $\Delta$ is given by
\begin{equation} \label{deltaati}
\Delta_{mn} = I_p+U_p+p^2/2-(n+2m)\hbar\omega_0\,.
\end{equation}
From \eq{deltaati} and \eq{propa} follows that ATI peaks appear at integer 
multiples $n\hbar\omega_0$ of the laser frequency, when
\begin{equation} \label{nochmalatispek}
\frac{p^2}{2} \, = \, n \hbar \omega_0 - I_p - U_p \, .
\end{equation}
One can also see from \eq{propa} that the ATI maxima become sharper
with each optical cycle that supplies ionizing trajectories. 
Of course, this effect is weakened by  the spreading of the wavepacket
hidden in the prefactor of each trajectory contribution (see \eq{hkpsi})
 not considered here.
 
Trajectories that are ionized during different laser cycles accumulate
a specific mean phase difference.  The phase difference depends on the
number $k$ of laser cycles passed  between the two
ionization processes:
\begin{equation} \label{deltaphiati}
\Delta \Phi(p) \, = \, k \, T \, \left( I_p + \frac{p^2}{2} + U_p \right) \, .
\end{equation}
The trajectories interfere constructively if
\begin{equation} 
\Delta \Phi(p) \, = \, 2 \pi l \quad \Rightarrow \quad \frac{1}{2} \, p^2 \, = \, 
\frac{l}{k} \omega_0  - I_p - U_p \label{atirational} \, .
\end{equation} 
If an energy spectrum is calculated exclusively with trajectories from
two intervals separated by $k$ cycles there should be additional
maxima in the ATI spectrum with a distance $\hbar \omega_0/k$.

\begin{figure}
\pic{8}{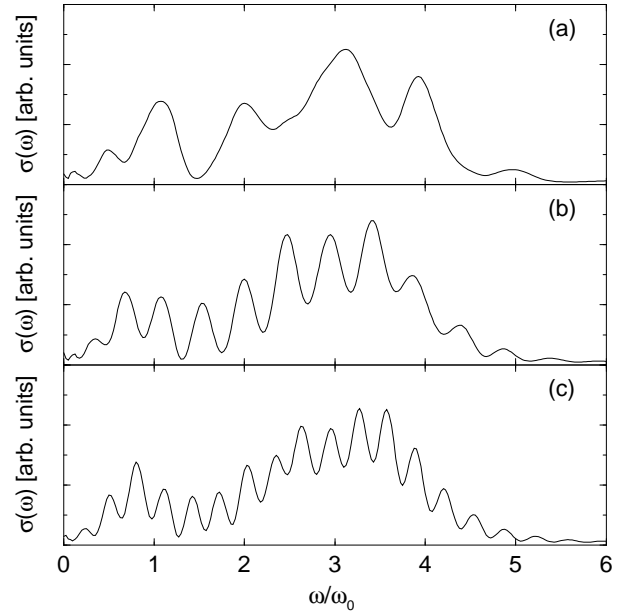}
\caption[]{Semiclassical spectra calculated with trajectories from the intervals 
$I_2$ and $I_4$ (a), $I_2$ and $I_6$ (b), and $I_2$ and $I_8$ (c).} 
\label{atihalbe}
\end{figure}

As a test for this semiclassical interpretation of the ATI 
mechanism we have calculated three
spectra with trajectories where the mean time delay between ionizing events
is given by $\Delta t = T$, $\Delta t = 2T$ and $\Delta t = 3T$.  For
the spectrum \fig{atihalbe}\!  (a) we have used exclusively
trajectories from the intervals $I_2$ and $I_4$ ($\Delta t = T$). 
One can see broad maxima separated by $\hbar \omega_0$
in energy.  Trajectories from the intervals $I_2$ and $I_6$ (see
\fig{atihalbe}\!  (b)) form a spectrum where the maxima are separated
by $\hbar \omega_0/2$ -- as predicted for $\Delta t = 2T$.  In analogy
the separation  for the ATI maxima in a spectrum with trajectories from
the intervals $I_2$ and $I_8$ is given by $\hbar \omega_0/3$ (Abb.\
\ref{atihalbe}\!  (c)).  The interference of trajectories ionized in
many subsequent cycles suppresses the non-integer maxima according to
\eq{deltaati}.  If the field strength is high enough the atom is
completely ionized during the first cycle.  The opportunity for
interference gets lost and we end up with an unstructured energy
spectrum. 

In an extreme semiclassical approximation we would have evaluated
the integral in \eq{propa} by stationary phase. The condition
\begin{equation} \label{diffATI}
d/d\tau[\Phi_b(\tau)-\Phi_c(\tau)] \equiv I_p+ p^2(\tau)/2 = 0
\end{equation}
leads to complex ionization times $t_n$ whose real part is periodic
and allows for two ionizing events per laser cycle, close to the extrema of the 
laser amplitude. The derivation is simple but technical, therefore we
don't carry it out explicitely here. However, it explains the observation that
ionization occurs close to the extrema of the laser field and it also makes
contact with the tunnelling process often referred to in the literature since
the complex time can be interpreted as  tunnelling at a complex
"transition" energy.

Clearly, our semiclassical analysis as described here supports the picture which 
has been sketched in \cite{LK97} interpreting a quantum calculation.
The authors assume that wave packets are emitted every time the laser reaches an extremum. 
The interference of the different wave packets gives rise to the ATI 
peaks.

In the following we will discuss the process of higher harmonic
generation (HHG) which is closely related to ATI. In fact, the separation
into a bound and continuum part of the electron description is constitutive
for HHG as well, the prominent
features, such as cutoff and  peak locations, can be derived from the
same phase properties \eq{propa} as for ATI. However, there is a characteristic
difference, {\it how} these phases enter.

\section{High Harmonic Generation}
\begin{figure}
\pic{8}{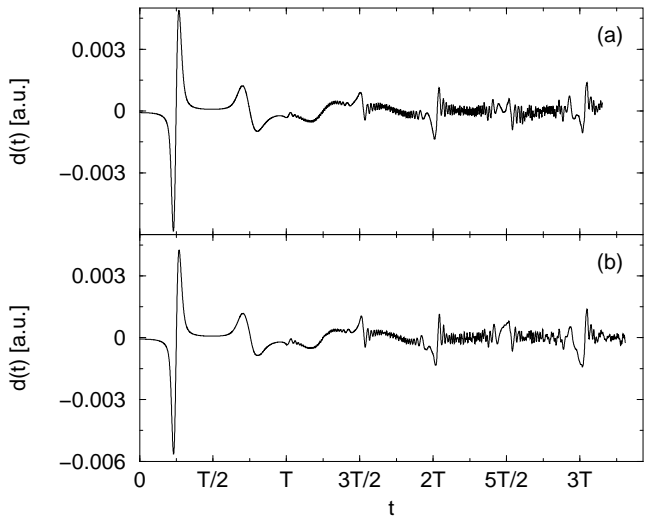}
\caption[]{Quantum (a) and semiclassical (b) 
dipole acceleration of higher harmonics according to \eq{dipol}.} \label{dipolqmscl}
\end{figure}

\begin{figure}
\pic{8}{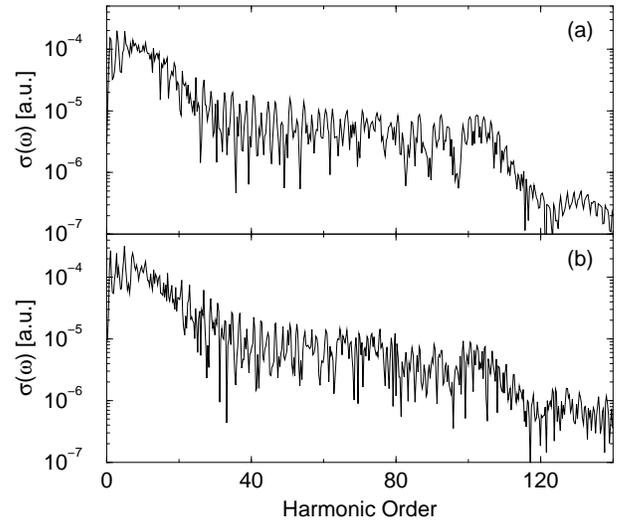}
\caption[]{Quantum (a) and semiclassical (b) 
spectrum of higher harmonics according to \eq{spec}.} \label{spectrum}
\end{figure}

\begin{figure}
\pic{8}{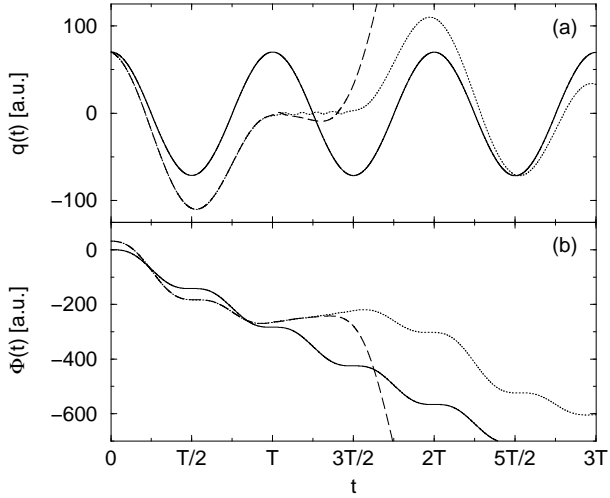}
\caption[]{Examples for direct (solid line), trapped (dotted line), and stranded (dashed line)
trajectories, see text.}
\label{trajek}
\end{figure}

First, we briefly recapitulate the findings of \cite{hhg99}, where 
we have calculated the harmonic spectrum
with the softcore potential  \eq{soft}.  With  our choice of $a=2$
the ionization potential is given by $I_p = 0.5$\! a.u..
The laser field has a strength $E_0 = 0.1$\! a.u., a frequency $\omega_0 = 0.0378$\! a.u.\ 
and a phase $\delta = \pi/2$. The initial wave packet with a 
width of $\beta = 0.05$\! a.u.\ is located at 
$q_\beta = E_0 / \omega_0^2 = 70$\! a.u.. Note, that the cutoff energy $E_C$
in such a symmetric laser scattering experiment is given by 
\begin{equation}
E_C \, = \, I_p + 2 U_p \, .
\end{equation}
From the dipole acceleration (see \fig{dipolqmscl})
\begin{equation} \label{dipol}
d(t) = - \left < \Psi(t) \left | \frac{dV(x)}{dx} \right | \Psi(t) \right > \, ,
\end{equation}
follows by Fourier transform 
\begin{equation} \label{spec}
\sigma(\omega) = \int \! d(t) \, \exp ( i\omega t)  \, dt
\end{equation}
the harmonic power spectrum (see \fig{spectrum}).
Clearly, our 
semiclassical approach represents a good approximation.
The dipole acceleration shows the characteristic feature that fast oscillations
(which are responsible for the high harmonics in Fourier space) show only
up after some time, here after $t = T$. 
This is the first time where trajectories are trapped. Trapping can only occur if 
(i) $t_n = n T/2$, (ii) the trajectories reach a turning point (i.e. $p(t_n) = 0$), and (iii)
if at this time the electron is close to the nucleus ($q(t_n)\approx 0$).
%
%
The trapped trajectories constitute a partially bound state which 
can interfere with the main part of the wavepacket (trajectories)
still bouncing back and forward over the nucleus driven by the laser.
The group of briefly bound (i.e.\ trapped or stranded) trajectories can be clearly identified,
either by their small excursion in space (\fig{trajek}a) or by the positive
slope of their action (\fig{trajek}b) as it was the case for ATI (compare
with \fig{int246}). By artificially discarding the initial conditions
in the semiclassical propagator which lead to trapped trajectories 
one can convincingly demonstrate that the plateau in HHG generation is
a simple interference effect \cite{hhg99}. Here, we are interested firstly
in linking ATI to HHG by using the same  separation in bound and
continuum parts of the dynamics already worked out for ATI. Secondly,
we want to go one step further and construct a wavefunction based
on this principle.

\begin{figure}
\pic{8}{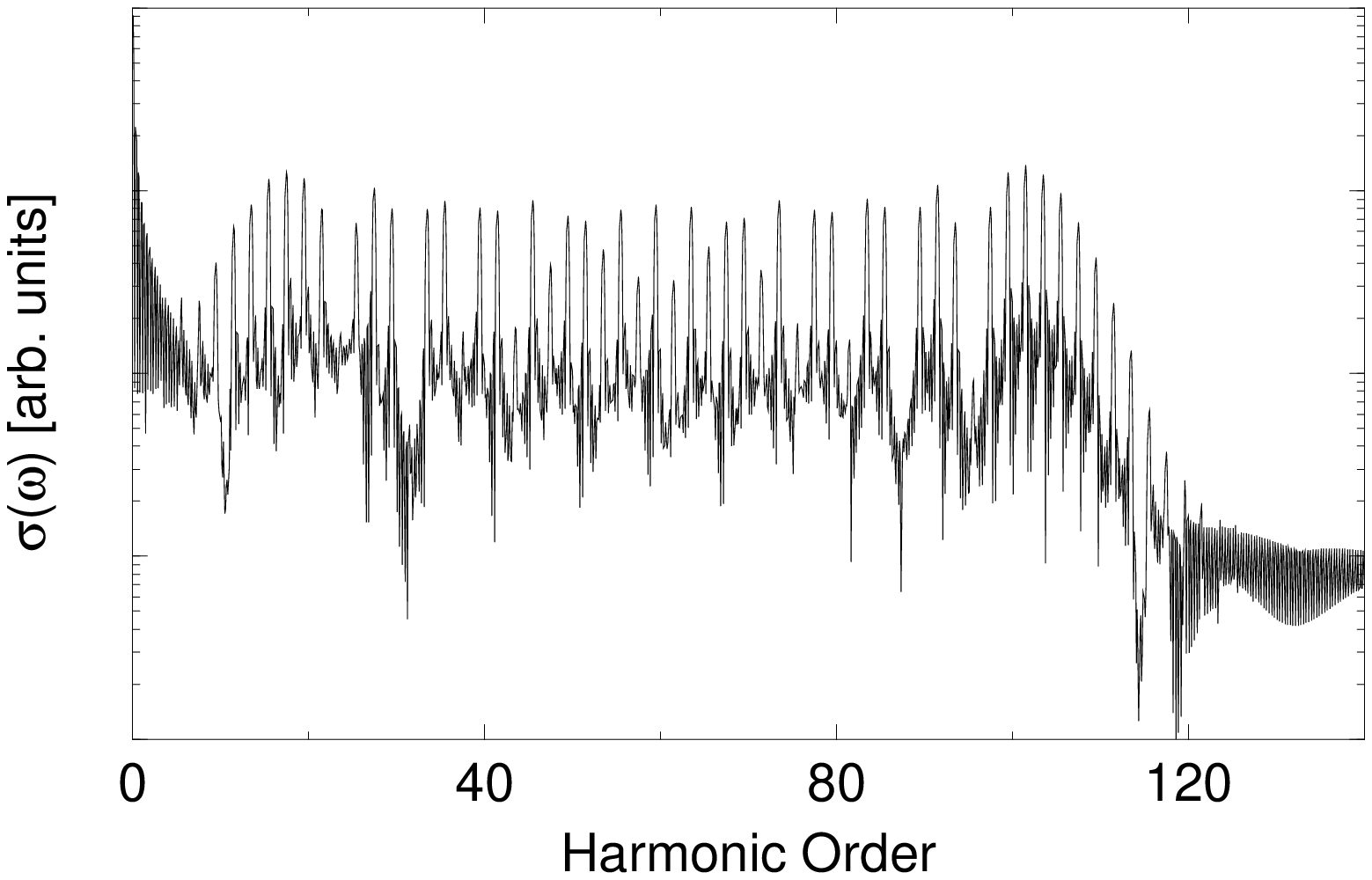}
\caption[]{Harmonic spectrum according to \eq{annaint}.}
\label{anna}
\end{figure}

Semiclassically, we have to look first at the phases of the 
observable. Therefore, we define a linear combination
for the wavefunction from the respective phase factors for bound and continuum
motion.   Considering only terms in the exponent
the harmonic spectrum \eq{spec} reads simply
\begin{align} \label{annaint}
& \sigma(\omega) \, \sim  \, \int \! \! dt \, \, \exp(i \omega t) \nonumber \\
& \quad  \left| \exp \left( i\Phi_c(t)/\hbar \right)
+ c\exp \left( i\Phi_b(t)/\hbar \right) \right |^2\,,
\end{align}
where $c\ne 0$ is a (so far) arbitrary constant. In principle, $c = c(t)$,
however its  change in time is much slower than that of the optical oscillations of
the phases $\Phi(t)$, hence we may approximate $c$ by a constant. 
The bound and continuum phases, $\Phi_b$ and $\Phi_c$, are 
defined in \eq{baction} and \eq{freewirkneu}, respectively. For $\Phi_c$ 
we have $p=0$, since this is the dominant contribution
from the center of the  wavepacket which was initially at rest.
The result is shown in \fig{anna}. Indeed, the plateau with the harmonics
is generated, however, the initial exponential decrease is missing
since we have neglected all prefactors of the semiclassical wavefunction which
describe the dispersion of the wavepacket.

Consequently, one can evaluate \eq{annaint} in stationary phase approximation.
The integrand of \eq{annaint} becomes stationary if 
\begin{equation} \label{diffHHG}
\frac{d}{dt} \left[ \hbar\omega t \pm 
\left(\Phi_b(t)-\Phi_c(t) \right) \right] \, = \, 0 \, 
\end{equation}
which happens at 
\begin{equation}\label{cutoff}
\hbar \omega = 2U_p \sin^2(\omega t) + I_p\,.
\end{equation}
From \eq{cutoff} we conclude the cut-off law 
\begin{equation}
\omega_{\textrm{max}} = 2 U_p + I_p \, ,
\end{equation}
as expected for laser assisted  electron ion scattering \cite{hhg99}.
Using the same expansion into Bessel functions as in \eq{propa} we obtain 
for the spectrum  
 \eq{annaint}:  
\begin{align}
& \int \! \! dt \, \, 
\exp \left( \frac{i}{\hbar} \left[ \left( \hbar \omega - U_p - I_p \right) t + \frac{U_p}{2 \omega_0} \sin 
\left(2 \omega_0 t \right) \right] \right) \nonumber \\
& = \, \sum \limits_{k=-\infty}^{\infty} 
\int \! \! dt \, \, e^{it \left( \hbar \omega - U_p - I_p + 
2 k \hbar \omega_0 \right)/\hbar}  \,  \, 
\textrm{J}_k \left(\frac{U_p}{2 \hbar \omega_0} \right) \, .
\label{hhganabessel}
\end{align}
Therefore, we see 
 maxima in the harmonic spectrum for
\begin{equation} \label{hhganaomega}
\hbar \omega_k \, = \, U_p + I_p - 2k \omega_0 \, .
\end{equation}

\begin{figure}
\pic{8}{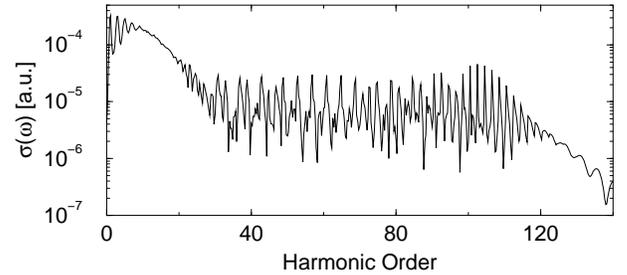}
\caption[]{Harmonic spectrum, generated from the wavefunction \eq{admix}
with $c = 0.025$ and $\beta = 0.05$ a.u..}
\label{harmonicana}
\end{figure}

We can go one step further and construct a full time-dependent 
wavefunction from this semiclassical approximation, namely
\begin{equation} \label{admix}
\Psi(x,t) \, = \, \Psi_\beta^{\scriptscriptstyle sc}(x,t) +
 c \, \Psi_0(x) \exp (it I_p /\hbar).
\end{equation} 
Here, $\Psi_0(x) \exp (i I_p t/\hbar)$ is the time dependent ground state
wave function (without the laser field) and $\Psi_\beta^{\scriptscriptstyle sc}(x,t)$
is a (semiclassical) {\it wavepacket} in the laser field but without potential.
Calculating the dipole acceleration  and the resulting
harmonic spectrum with this wavefunction leads to a remarkably good approximation of the true
quantum spectrum (compare \fig{spectrum} with \fig{harmonicana}). The dispersion
of the wavepacket leads to the lower plateau compared to \fig{anna}.
\section{Conclusions}
\subsection{Semiclassical comparison between ATI and HHG}
Clearly, the main structure  such as the plateau, cutoff (HHG) and the
 occurrence of peaks and
their separation in energy (ATI  and HHG) is a property of the difference
of the classical time-dependent actions $\Phi_b(t) -\Phi_c(t)$ alone.

However, the HHG power spectrum \eq{spec} is an integral over all the time for which 
 the electron wavepacket is exposed to the laser field. In contrast, the ATI
 spectrum is obtained in the long-time limit $t\to\infty$ after the laser 
 has been switched off. This difference may explain why the HHG results tend
 to be better than the ATI results semiclassically: Any semiclassical
 approximation (which is not exact) become worse for large times.
 
 
 A second point refers to the fact that the characteristic phase difference
 $\Phi_b(t) -\Phi_c(t)$ appears already in the wavefunction \eq{propa} 
 for ATI, while for HHG it occurs only in the expectation value \eq{dipol}.
 However, this difference is artificial, since the expectation value, or 
 better its Fourier transform the power spectrum, is not the observable
 of higher harmonic radiation. The correct expression is the dipole-dipole
 correlation function $R$ which can be approximated as $R\propto
 |\sigma(\omega)|^2$ under single atom conditions or in the case of an ensemble
 of independent atoms which radiate \cite{BLM94,SuMi90}. Hence, in both cases,
 ATI and HHG, the peak structure appears already on the level of the quantum 
 amplitude (or wavefunction) and is amplified in the true observable.

\subsection{Summary}
We have given a time-dependent fully semiclassical description of 
multiphoton processes. The prominent ATI and HHG features emerge naturally
from properties of the classical trajectories whose contributions to the
semiclassical wavefunction interfere semiclassically. Any effect of this
semiclassical interference
can be  double-checked by disregarding the phases. This leads (with the 
same trajectories) to a classical observable.
As we have seen, to a good approximation 
the classical action for an individual trajectory
can be composed of one part $\Phi_b$
for the time the electron is bound (disregarding the laser field) and 
of another part $\Phi_c$ for the time the electron is in the continuum
(disregarding the atomic potential). The relevant phase difference
$\Phi_b-\Phi_c$ leads in both cases, ATI and HHG, to the prominent harmonic
structures in terms of the laser energy $\hbar\omega_0$. Finally, we have been
able to construct a simple wavefunction for higher harmonics generated
in laser assisted scattering. Its key element
is an explicitely time-dependent wavepacket of the electron under the
influence of the laser field. Starting from an initial Gaussian distribution
localized in space the wavepacket disperses in time providing the correct
decrease of the intensity of the lower harmonics and in turn the correct
height of the plateau.

Financial support from the  DFG under the 
Gerhard Hess-Programm and the SFB 276 is  gratefully acknowledged.
\begin{appendix}
\section*{}
We want to calculate the semiclassical wave function of a free particle in a laser field
according to \eq{hkpsi}.
A particle in a laser field $V_L(x,t) = E_0 \sin(\omega t)$ moves with
\begin{align}
& p(t) \, = \, p + \frac{E_0}{\omega} \cos(\omega t) \, \equiv \, p + \tilde{p}(t)  
\\
& q(t) \, = \, q + p \, t + \frac{E_0}{\omega^2}  \sin(\omega t) 
       \, \equiv \, q + p \, t + \tilde{q}(t)  
\end{align}
The weight factor $C_{q p}(t)$ is given by
\begin{equation}
C_{q p}(t) \, = \, \left(1 - \frac{i \hbar \gamma}{2} t \right)^{\frac{1}{2}} \, .
\end{equation}
For the phase factor $S_{q p}(t) - p(t) q(t)$ we get: 
\begin{align}
S_{qp}(t) - p(t) q(t) \, = \, & -\frac{U_p}{2 \omega} \sin(2 \omega t) - U_p \, t 
\nonumber \\
& \,  - \frac{p^2}{2} t - \tilde q(t) \, p - q \, p 
\end{align}
Evaluating \eq{hkpsi} with 
the stationary phase approximation, which is exact for quadratic 
potentials, leads to the condition that 
\begin{align} 
f(q,p) \, = \, & \frac{i}{\hbar} \left( x \, p(t) - \frac{p^2}{2} t - \tilde{q}(t) \, p - 
\frac{\gamma}{\alpha} q \, p - \frac{\beta}{\alpha} q_\beta  \, p \right) 
\nonumber \\
& \, - \frac{\gamma}{2} \left( x - q(t) \right)^2 - \frac{\gamma \beta}{2 \alpha}  
\left( q - q_\beta \right)^2 - \frac{1}{2 \hbar^2 \alpha} p^2  
\end{align}
must have an extremum. With   
\begin{align}
& \frac{\partial f}{\partial q} \, = \, 0 \, = \, \gamma \left[ x - q(t) \right] - 
\frac{\gamma \beta}{\alpha} (q - q_\beta) - \frac{i}{\hbar} \frac{\gamma}{\alpha} p 
\\[0.2cm]
& \frac{\partial f}{\partial p} \, = \, 0 \, = \, \gamma \left[ x - q(t) \right] t - 
\frac{1}{\hbar^2 \alpha} p \nonumber \\
& \qquad \qquad \quad + \frac{i}{\hbar} \left( x - p \, t - \tilde q(t) - 
\frac{\gamma}{\alpha} q - \frac{\beta}{\alpha} q_\beta \right) 
\end{align}
we find
\begin{align}
& q_s \, = \, \frac{x - \tilde q(t) + i \hbar \beta t q_\beta}{1 + i \hbar \beta t} 
\\
& p_s \, = \, \frac{i \hbar \beta}{1 + i \hbar \beta t} \left(x - \tilde q(t) - q_\beta \right) \, .
\end{align}
After some algebra we arrive at  the stationary exponent 
\begin{align}
f(q_s,p_s) \, &  = \, \frac{i}{\hbar} x \, \tilde p(t) - \frac{\beta}{2 \left(1 + i \hbar \beta t \right) } 
\left( x - \tilde q(t) - q_\beta \right)^2
\nonumber \\[0.2cm]
 & = \, \frac{i}{\hbar} x \, \tilde p(t) 
- \frac{i}{\hbar} \frac {\hbar^2 \beta^2 t}{2 \sigma(t)} \left( x - \tilde q(t) - q_\beta \right)^2 
\nonumber \\
& \quad - \frac{\beta}{2 \sigma(t)} \left( x - \tilde q(t) - q_\beta \right)^2 \, ,
\end{align} 
where $\sigma(t)$ is given by
\begin{equation}
\sigma(t) \, = \, 1 + \beta^2 \hbar^2 t^2 \, .
\end{equation}
The determinant of the second derivatives of $f$ still has to be calculated.
With 
\begin{align} 
& \frac{\partial^2 f}{\partial q^2} \, = \, - \frac{\gamma^4 + 2 \gamma \beta}{\alpha} \qquad \quad
\frac{\partial^2 f}{\partial p^2} \, = \, - \frac{i}{\hbar} t - \gamma t^2 - \frac{1}{\hbar^2 \alpha}  
\nonumber \\[0.2cm]
& \frac{\partial^2 f}{\partial q \partial p} \, = \, - \frac{i}{\hbar} 
\frac{\gamma}{\alpha} - \gamma t 
\end{align}
we get
\begin{equation}
\det \begin{pmatrix}
\displaystyle{\frac{\partial^2 f}{\partial q^2}}  & 
\displaystyle{\frac{\partial^2 f}{\partial q \, \partial p}} \\[0.4cm]
\displaystyle{\frac{\partial^2 f}{\partial p \, \partial q}}  & 
\displaystyle{\frac{\partial^2 f}{\partial p^2}}
\end{pmatrix} \, = \, \frac{2 \gamma}{\hbar^2 \alpha} 
\Bigl( \left[ 1 - i \gamma \hbar t / 2\right] \left[ 1 + i \beta \hbar t \right] \Bigr) \, .
\end{equation}
The factor $\gamma$ cancels as it should be and we are left with
\begin{align} 
\Psi_\beta^{\scriptscriptstyle sc}(x,t) \,   = \, & \left( \frac{\beta}{\pi} \right)^{1/4} 
\sqrt{\frac{1}{1 + i \hbar \beta t}} 
\nonumber \\
& \, \exp \left( \frac{i}{\hbar} \left[\tilde{p}(t) \, x -\frac{U_p}{2 \omega} \sin(2 \omega t) - U_p t
  \right] \right) 
\nonumber \\
& \, \exp \left( \frac{i}{\hbar} \frac{\hbar^2 \beta^2}{2 \sigma(t)} \left( x - \tilde{q}(t) - q_\beta \right)^2 
t \right) 
\nonumber \\
& \, \exp \left( - \frac{\beta}{2 \sigma(t)} \left( x - \tilde{q}(t) - q_\beta \right)^2 \right) \, .
\end{align}
This semiclassical time dependent wavepacket is 
quantum mechanically exact and corresponds to a
superposition of Volkov solutions according to a Gaussian distribution at
time $t=0$ \cite{VTP99}.  The fact that the semiclassical 
wavefunction is exact is a direct consequence of the Ehrenfest theorem 
which implies that interactions $V\propto x^n$, $n = 0,1,2$ have 
quantum mechanically exact semiclassical solutions. 
\end{appendix}

\vspace*{-3mm}



\begin{references}
\bibitem{rem} As already discussed in \cite{JES88} it is therfore important that the
calculatuions are stopped at times $t_m=(2m-1)T/4$ so that the ionized
trajectories terminate at a time where they have their mean momentum.
\end{references}
\end{document}